\documentclass[aps,prl,a4paper,reprint,showpacs,amsmath,amssymb,superscriptaddress,floatfix]{revtex4-1}

\usepackage{graphicx}

\begin{document}

\title{Transport in helical Luttinger Liquid with Kondo impurities}

\author{Oleg M. Yevtushenko}
\affiliation{Ludwig Maximilians University, Arnold Sommerfeld Center and Center for Nano-Science, Munich, DE-80333, Germany}

\author{Ari Wugalter}
\affiliation{Department of Physics and Astronomy, Rutgers The State University of New Jersey,
                 136 Frelinghuysen rd., Piscataway, 08854 New Jersey, USA}

\author{Vladimir I. Yudson}
\affiliation{Institute of Spectroscopy, Russian Academy of Sciences, 5
                 Fizicheskaya Street, Troitsk, Moscow 142190, Russia}
\affiliation{Russian Quantum Center, Skolkovo, Moscow Region 143025, Russia}

\author{Boris L. Altshuler}
\affiliation{Physics Department, Columbia University, 538 West 120th Street, New York, New York 10027, USA}

\begin{abstract}
Ballistic transport of helical edge modes in two-dimensional topological insulators is
protected by time-reversal symmetry. Recently it was pointed out \cite{AAY} that coupling of
non-interacting helical electrons to an array of randomly anisotropic Kondo impurities
can lead to a spontaneous breaking of the symmetry and, thus, can remove this protection.
We have analyzed effects of the interaction between the electrons using a combination
of the functional and the Abelian bosonization approaches. The suppression of the ballistic
transport turns out to be robust in a broad range of the interaction strength. We have
evaluated the renormalization of the localization length and have found that, for strong
interaction, it is substantial. We have identified various regimes of the dc transport
and discussed its temperature and sample size dependencies in each of the regimes.
\end{abstract}

\date{\today}


\maketitle

{\it Introduction}:
Electron transport in time-reversal invariant topological insulators (TI)
%%%
% low-dimensional  helical systems, which exist at boundaries
% of time-reversal invariant topological insulators,
%%%
has become a hot topic of research during past several years,
see Refs.\cite{HasanKane,QiZhang,TI-Shen,MolFranz} for reviews.
The bulk electron states in the TIs are gapped, nevertheless, dc transport is
possible since it is provided by low-dimensional  helical edge modes.
Helicity means the lock-in relation between electron spin and momentum:
%%%
% caused by strong spin-orbital coupling.
%%%
helical electrons propagating in opposite directions have opposite spins \cite{WuBernevigZhang,XuMoore}.
%%%
% ; one may somewhat conventionally speak about right (left) movers with spin up (down), respectively.
%%%
% Time-reversal symmetry ensures equal energy of electron states with opposite momentum
% and spin, $ \, | k, \uparrow \rangle \, $ and $ \, | -k, \downarrow \rangle $.
%%%
% ($ \, k \, $ is the electron momentum, the arrow indicates the spin direction).
%%%
% When the time-reversal symmetry is not broken, the electron states $(k, \uparrow)$
% and $(-k, \downarrow)$ are of the same energy ($k$ is electron momentum,
% the arrow indicates the spin direction), thus
An elastic backscattering of a helical electron must be accompanied by a spin-flip.
Therefore, the helical electrons are immune to effects of potential disorder such as
localization.
%%%
% The ballistic dc conductance of the helical system seems to be protected since it is insensitive to the disorder strengths.
%%%
% That is why the TIs with the helical edges are even referred to as ``a new state of matter'' \cite{?}.
%%%
% This protection of the ballistic transport has resulted in a huge interest in one-dimensional (1D)
% helical edge modes, which is partially stimulated by a hope to develop new spintronic and electronic
% devices.
%%%

Recent experimental studies of the charge transport through 1D channels
at the helical edges of 2D TIs
\cite{Molenkamp-2007,EdgeTransport-Exp0,BismuthTI-Meas,InAs-GaSb-Meas}
made of quantum-wells \cite{BHZ,AssymQ-Wells} demonstrated that the
transport is indeed close to be ballistic as long as the samples are small.
\cite{Molenkamp-2007,Molenkamp-2009}.
However, longer edges exhibit lower conductance
\cite{Molenkamp-2007,EdgeTransport-Exp1,EdgeTransport-Exp0}
which is evidence for back-scattering.
%%%
% , moreover, the edge resistance can be T-independent in a broad ranges of temperatures \cite{EdgeTransport-Exp2}.
%%%
% Moreover, since the sub-ballistic conductance can be temperature-independent
% in a broad range of temperatures \cite{EdgeTransport-Exp2} this back-scattering
% might be elastic.
%%%
Moreover, the absence of clear temperature dependence of the sub-ballistic conductance
in a broad temperature interval \cite{EdgeTransport-Exp2} suggests that this back-scattering
is probably elastic.

Robustness of the ballistic transport in the Tis was discussed in several theoretical papers.
%%%
% A number of theoretical investigations has addressed robustness of the ballistic transport in the TIs.
%%%
% where either the edge disorder has spin degrees of freedom or the helical electrons interact.
%%%
% One of the search directions is accounting for impurities with spin degrees of freedom.
%%%
Spin-flips needed for the backscattering could be, for example, due to spin-1/2
Kondo impurities. However, the Kondo-screening of the spin would recover the ballistic
transport at low temperatures \cite{MaciejkoOregZhang,FurusakiMatveev}.
The inelastic processes caused by interactions in the presence of disorder are
predicted to result in temperature-dependent contributions to the dc conductance
and conductance fluctuations \cite{FvOpp-Inelast,CheiGlaz,GoldstGlazman-Short,GoldstGlazman-Long,Mirlin-HLL}.
Such corrections are also frozen out at cooling due to the lack of the
phase space.
%%%
% At the same time, low temperature experiments reveal a suppression of the ballistic
% dc transport \cite{EdgeTransport-Exp1},
%%%
% For example, the edge transport in long topological insulators made of  InAs/GaSb
% quantum wells  shows that the effective edge resistance is much greater than $ \, h/e^2 \, $
% in a temperature from 20 mK to 1K \cite{EdgeTransport-Exp1}; this large resistance
% can be temperature independent in a broad range of temperatures \cite{EdgeTransport-Exp2}.
%%%
Thus, the explanation of the dc transport in the TIs, which (i) is non-ballistic at low temperatures,
and (ii) can be temperature independent, remains a theoretical challenge.
%%%
% Thus, a theory which would explain the low-temperature backscattering of helical edge
% electrons and describe the temperature dependence of conductance is still in demand.
%%%

The first step toward the understanding of the non-ballistic temperature-independent
transport through 1D helical edges was taken in Ref.\cite{AAY} where the
idea of spontaneous breaking of time-reversal symmetry was proposed. It was demonstrated
that helical 1D electrons, which do not interact with each other, are localized if they are
coupled to an array of the Kondo impurities. This coupling can be described by the Hamiltonian:
\begin{equation}
\label{Hb}
      \hat{H}_{b} = \int \!\! {\rm d} x \, \rho_s J_{\perp}
       \left[ ( S^+ + \epsilon \, S^- ) e^{2 i k_F x} \psi^{\dagger}_{-} \psi_{+} + h.c. \right] \!\! .
\end{equation}
Here $ \, \psi_+ $ \, ($\psi_-$) describes spin up right moving (spin down left moving) in
$x$-direction helical fermions $ \, \psi_{R,\uparrow} \, $ ($\psi_{L,\downarrow}$);
$ k_F \, $ is the Fermi momentum; $ \, \rho_s \, $ is the impurity density;
$ \, J_{\perp} \equiv ( J_x + J_y ) / 2 \, $ is the coupling constants between the Kondo-
and the fermion- spins; $ \, \epsilon \equiv ( J_x - J_y ) / J_{\perp} \, $ is the dimensionless
parameter of the anisotropy in the plane of the 2D TI (XY-plane); $ \, S^{\pm} \, $ are the Kondo
spin operators.

For isotropic couplings, $ \, \epsilon = 0 $, the indirect interaction between spins induces
a slowly varying in space and time spin polarization. Homogeneous time-independent polarization
would create a gap, $ \, \Delta_0 = s \rho_s J_{\perp} $, in the spectrum of fermions
($ s = 1/2 \, $ is the impurity spin). Fluctuations of the polarization result in heavy but
gapless ``polaronic'' complexes of helical electrons dressed by slow spinons. These complexes
are charged and can support ballistic transport with a strongly reduced Drude weight. A random
anisotropy, $ \, \epsilon(x) \ne 0 $, quenches the local spin polarization and thus causes
spontaneous symmetry breaking. The polaronic complexes loose their protection from the back-scattering
and undergo the Anderson localization with maybe large but finite localization radius,
$ \, {L}^{(0)}_{\rm loc} $.

\begin{figure}[t]
   \includegraphics[width=0.475 \textwidth]{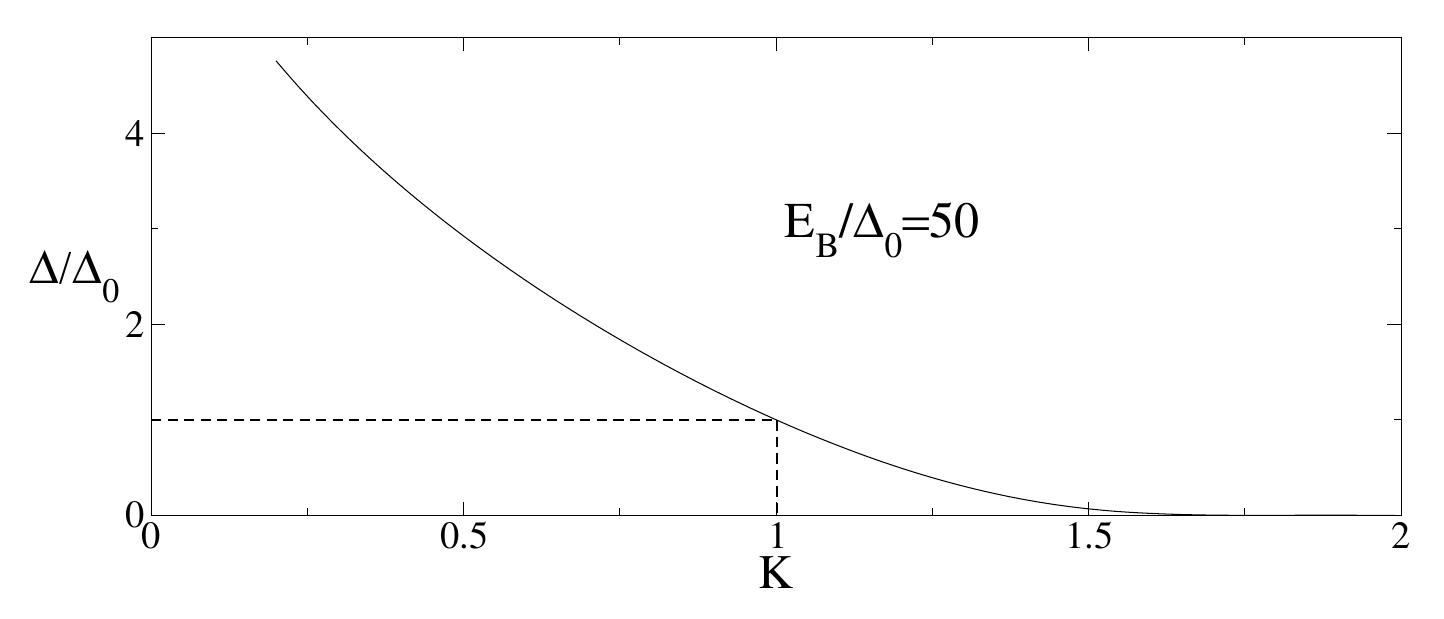}
   \caption{\label{Delta_K}
        Typical dependence of the effective gap on the Luttinger parameter,
        see Eq.(\ref{Delta-L-Eff}).
%%%
% The inset ADD? shows a comparison of $ \, \Delta \, $ with the
%        Kondo temperature of the single impurity coupled to the HLL.
%%%
                }
\end{figure}

{\it In this Letter}, we explore the charge localization in the system of interacting helical
electrons - Helical Luttinger Liquid (HLL) - coupled to the array of the Kondo impurities, and
analyze the temperature dependence of the dc transport.
Interactions in the HLL are characterized by the Luttinger parameter
$ \, K = (1 + g)^{-1/2} $ ($ \, g \, $ is the dimensionless interaction strength).
We prove that the moderate attraction, $ \, 1 < K < 2 $, and (almost) arbitrary repulsion,
$ \, K < 1 $, do not change the qualitative picture of the non-interacting system:
the effective theory, which describes localization in the HLL coupled to the Kondo array,
remains valid, though the gap, $ \, {\Delta}$, and the localization radius, $ \, L_{\rm loc} $,
are substantially renormalized:
\begin{equation}
\label{Delta-L-Eff}
   \frac{{\Delta}(K)}{\Delta_0} \sim K \left( \frac{E_B}{K^2 \Delta_0} \right)^{ \frac{1-K}{2-K} } \!\!\! ; \
%%%
% \!\!\! , \  \Delta_0 = \Delta(K=1) \, ; \\
%%%
   \frac{{L}_{\rm loc}}{{L}_{\rm loc}^{(0)}} \sim
%%%
%                                               \left( \frac{a \rho_s}{w} \right)^{1/3}
%%%
                                               \left( \frac{ \Delta_0 }{ K \, {\Delta}(K) } \right)^{\frac{4}{3}} \!\!\! ;
%%%
% {\cal L}_{\rm loc}^{(0)} = {\cal L}_{\rm loc}(K=1) \, .
%%%
\end{equation}
%%%
% here $ \, \Delta_0 = s \rho_s J_{\perp} \, $ is the fermionic gap in the absence of the interactions;
%%%
% $ \, \rho_s \, $ is the average density of the Kondo impurities;
%%%
% $ \, s = 1/2 \, $ is the spin.
%%%
see Fig.\ref{Delta_K}. Here $ \, E_B \, $ is the UV energy cut-off which is of the order of the bulk gap in the TI.
%%%
% We restrict ourselves to low temperatures, $ \, T \le {\Delta} $, and do not consider inelastic processes cause by the
% electron interactions, cf. Refs\cite{GoldstGlazman-Short,GoldstGlazman-Long}.
% Other important assumptions are given in Eq.(\ref{Params}) and discussed in Sect. ``{\it Validity}''.
%%%

\begin{figure}[b]
   \includegraphics[width=0.475 \textwidth]{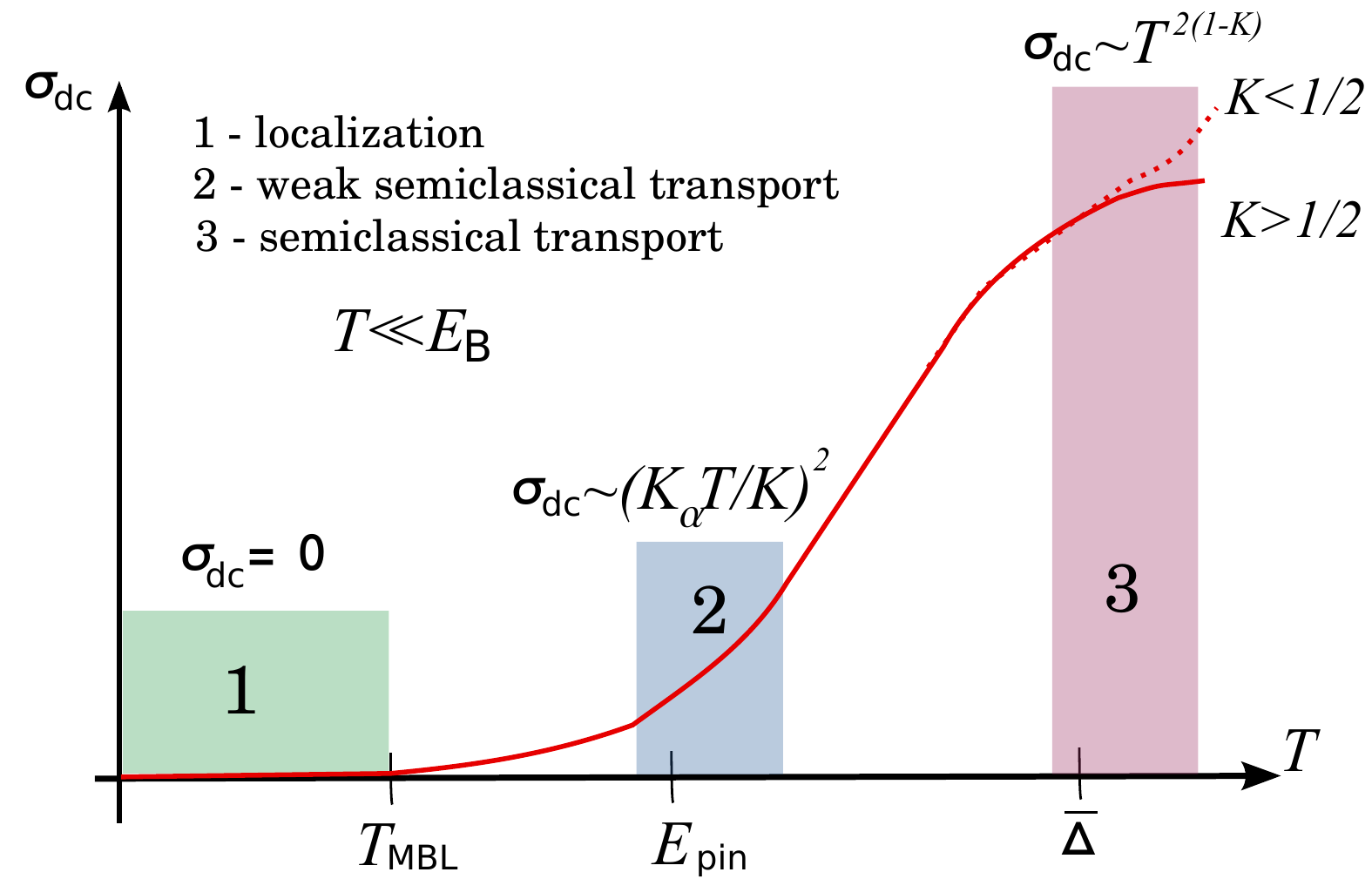}
   \caption{\label{SigmaDC_T}
        (color on-line)
        Sketch of the temperature dependence of the dc conductivity in different regimes, see explanations in
        Sect. ``{\it DC transport}'' and Eqs.(\ref{Sigma-C},\ref{Sigma-F}) in the text.
                 }
\end{figure}

The energy scales which govern different temperature regimes of the
dc transport are sketched on Fig.\ref{SigmaDC_T}. They are the temperature of many-body localization
transition \cite{MBL}, $ \, T_{\rm MBL} $, and the de-pinning energy, $ \, E_{\rm pin} $, defined below,
see Eq.(\ref{Epin}). We will assume that [see the discussion after Eq.(\ref{Epin})]
\begin{equation}
   T_{\rm MBL} \ll E_{\rm pin} \ll {\Delta} \, .
\end{equation}

If $ \, T <  T_{\rm MBL} \, $ all excitations are localized and ballistic transport is
suppressed. If $ \, T_{\rm MBL} < T < E_{\rm pin} $,
%%%
% many-body states of the composite particles are delocalized and
%%%
the dc conductivity is finite albeit low and is of a quantum
nature. The transport becomes thermally activated as $ \, T \to E_{\rm pin} $ and
turns into semiclassical one in the interval $ \, E_{\rm pin} < T < {\Delta} $.
%%%
% Transport of spinons is semi-classical at $ \, E_{\rm pin} < T < \Delta $.
%%%
% Helical fermions can contribute to dc transport when thermal fluctuations become of the order of the gap,
% $ \, T \to \Delta $.
%%%
Power-low dependencies of $ \, \sigma_{\rm dc} \, $ at $ \, T \gg T_{\rm MBL} \, $ result
from the interaction-dependent renormalization of $ \, J_{\perp} $, see Eqs.(\ref{Sigma-C},\ref{Sigma-F}).

{\it The model and calculations}:
The Hamiltonian of the HLL coupled to the array of the Kondo impurities is:
$ \, \hat{H} = \hat{H}_0 +  \hat{H}_{\rm int} + \hat{H}_b $, where
the first two terms describe the free fermions and the interaction between
them, respectively:
\begin{eqnarray}
\label{H0}
       \hat{H}_0 & = & - i v_F \int {\rm d} x \
                                       \sum_{\eta = \pm} \eta \, \psi^{\dagger}_\eta(x) \partial_x \psi_\eta (x) , \\
\label{Hint}
       \hat{H}_{\rm int} & = & \frac{g}{2 \nu} \int {\rm d} x \, \left( \rho_+ + \rho_- \right)^2 \!\! ,
       \quad
       \rho_\pm \equiv \psi^{\dagger}_{\pm} \psi_{\pm} ;
\end{eqnarray}
here $ \, v_F $ is the Fermi velocity and $ \, \nu \, $ is the density of states in the HLL. The
backscattering term $ \, \hat{H}_{b} \, $ (caused by the coupling of the electrons to the Kondo
impurities) is defined in Eq.(\ref{Hb}). We neglect the forward-scattering term $ \,
\sim J_z S_z \, $ since a unitary transformation of the Hamiltonian allows one to map the
model with the parameters $ \, \{ K, J_z \ne 0 \} \, $ to its counterpart
%%%
% with $ \, \{ \tilde{K}, J_z = 0 \} $
%%%
with the effective Luttinger parameter $ \, \tilde{K} = K ( 1 - J_z \nu / 2 K )^2 $ and $ \, \tilde{J}_z = 0 $
\cite{EmKivRes,MaciejkoLattice}. Thus, $ \, H_{\rm int} \, $ takes into account both the direct
electron-electron interaction and the interaction mediated by the z-coupling to the Kondo impurities.

Let us briefly recapitulate important steps, equations and basic assumptions from the paper
\cite{AAY}. The spin ordering allows us to develop a path integral formulation of the problem
using the parametrization of each spin by its azimuthal angle, $ \, \alpha $, and
projection on $z$-axis, $ \, n_z $:
\begin{equation}
    S^{\pm}(x,\tau) = s e^{i \alpha(x,\tau)} \sqrt{1 - n^2_z(x,\tau)} \, , \ |n_z| \le 1 \, ;
\end{equation}
$ \tau \, $ denotes the imaginary time. It is convenient to use the spins rotation:
$ \, S^{\pm}(x,\tau) e^{2 i k_F x} \to  S^{\pm}(x,\tau) $ which leads to the redefinition of
the anisotropy parameter: $ \, \epsilon e^{2 i k_F x} \to \epsilon e^{4 i k_F x} $.
%%%
% We are not interested in effects of incommensurability and assume that the impurities are
% located at (almost) all lattice sites such that the effective model for the anisotropy
% can be utilized: 1) if the Kondo array and the anisotropy are regular than $ \, e^{4 i
% k_F x} = 1 \, $ for all impurities and $ \,  \epsilon \, = {\rm const} $; 2) if the Kondo
% array and, correspondingly, the anisotropy are disordered and the disorder is
% smooth, we substitute  $ \, \epsilon(x) \, $
%%%
For a high density of the Kondo array with a weak irregularity in positions of the
spin impurities, the model can be simplified by treating $ \, \rho_s \, $ and $ \, \epsilon
e^{4ik_F x} \, $ as
%%%
% slowly varying in space
%%%
random variables: real $ \, \rho_s(x) \, $ and complex $ \, \epsilon(x) $.
%%%
% 1) If there is a weak irregularity of positions of the spin impurities
% but the anisotropy is regular than we put $ \, \epsilon e^{4 i k_F x} \to \epsilon = {\rm
% const} \, $ for all impurities and treat $ \rho_s \, $ as a smooth random function of $ \, x $;
% 2) If the irregularity of the Kondo array results in the random anisotropy the function
% $ \, \epsilon e^{4ik_F x} \, $ can be treated as a complex random variable $ \, \epsilon(x) $.
%%%
Fluctuations of $ \, \rho_s \, $ do not effect the dc transport in
the HLL, cf. Ref.\cite{AAY}, thus, one can use the averaged density. On the contrary,
randomness of $ \, \epsilon(x) \, $ plays the crucial role. Without loss of generality,
$ \, \epsilon(x) \, $ can be treated as a Gaussian random function with zero mean value
and short-range correlations
\begin{equation}
\label{EpsFluct}
   \langle \epsilon(x) \rangle_{\rm dis} = 0 \, ; \quad
   \langle \epsilon(x) \epsilon^*(x') \rangle_{\rm dis} = \frac{w}{\rho_s} \delta( x - x' ) \, .
\end{equation}
As a result, the Lagrangian density for the HLL coupled to the Kondo array can be presented as:
\begin{eqnarray}
\label{FermLagr}
   {\cal L} & = & \hat{\psi}^{\dagger} \left[
        \begin{array}{cc}
            \partial_+    &  {\cal J} \cr
            {\cal J}^* & \partial_-
        \end{array}
                                                \right] \psi + \frac{g}{2}  ( \hat{\psi}^{\dagger} \hat{\psi} )^2
                                                                  - i s \rho_s n_z \partial_\tau \alpha \, ; \\
        {\cal J} & \equiv & \Delta_0 \sqrt{1 - n_z^2} \, \left( e^{-i \alpha} + \epsilon e^{i \alpha} \right) \, ;
        \nonumber
\end{eqnarray}
where $ \, \partial_{\pm} = \partial_\tau \mp i v_F \partial_x \, $ are the chiral derivatives;
$ \, \hat{\psi}^{\dagger} \equiv \{ {\bar \psi}_+ , \,  {\bar \psi}_- \} \, $ is the fermionic
spinor field. Further calculations are based on the scale separation: the fermionic variables are
much faster than the spin ones. Therefore, as will be verified below,
\begin{equation}
\label{SlowAlpha}
   \xi \partial_x \alpha \ll 1 \, , \
   {\Delta}^{-1} \partial_\tau \alpha  \ll 1 \, ; \quad
   \xi \equiv v_F / {\Delta} \, .
\end{equation}

To describe the composite fermion-spinon excitations, we perform a gauge transformation of the
fermionic fields $ \, \psi_\eta e^{-i\eta\alpha/2} \to \psi_\eta $. After a point-split of the
interaction term and evaluation of the Jacobian, the Lagrangian density acquires the form:
\begin{equation}
\label{Lferm}
   \tilde{\cal L} [\psi, \alpha ] \simeq {\cal L}|_{ {\cal J}  \to  {\cal J} e^{i \alpha} } +
                 \frac{ v \left( \partial_x \alpha \right)^2}{8 \pi K} \, ;
%%%
%                 \frac{1}{4}{\cal L}_{\rm LL}(\alpha, K, v) \, ;
%%%
%   \frac{v_F}{8 \pi} (\partial_x \alpha)^2 \, ;
%%%
%   {\tilde{\cal J}} & \equiv & D \, \left( 1 + \epsilon e^{2 i \alpha} \right) \, .
%%%
\end{equation}
%%%
% see Ref.\cite{AAY} for details.
%%%
where $ \, v = v_F / K \, $ is the renormalized velocity, and sub-leading terms $ \, \rho_\eta \partial_\eta
\alpha \, $ are neglected in $ \, {\cal L} $ due to inequalities Eq.(\ref{SlowAlpha}).

We start our analysis of the effects of the electron-electron interaction with the zero anisotropy case,
$ \, \epsilon = 0 $. The classical configuration of the spin variables is
\[
  n_z^{\rm (cl)} = 0, \ \alpha^{\rm (cl)} = {\rm const}.
\]
Our goal is to derive an effective theory describing fluctuations around the classical solution. For $ \, g = 0 $,
this is achieved by integrating out all massive modes: firstly, the fermions with the gap $ \, \Delta_0 $,
and, secondly, the non-dynamical variable $ \, n_z $ (in the quadratic approximation).
The interaction can be taken into account with the help of the functional bosonization, which involves the
Hubbard-Stratonovich decoupling of the interaction term and the gauge transformation of the fermionic
fields \cite{Yurkevich-2001,Yurkevich-2004}. Following these standard steps, one can show that the main
non-perturbative effect of the weak interaction is the renormalization of the backscattering amplitude,
which acquires an effective  energy-dependence, $ \, J_{\perp} ({\cal E}) $. Similar to the bare relation
$ \, \Delta_0 \propto J_{\perp} $, we introduce the renormalized gap at each step of the renormalization
procedure: $ \, \Delta \propto J_{\perp}({\cal E}) $. Thus, the gap also becomes energy dependent. The
energy-dependence reads as:
\begin{equation}
\label{DeltaFB}
   \Delta({\cal E}) = \Delta_0 \times \left( \frac{E_B}{{\cal E}} \right)^{1-K} \!\! . % \quad | 1 - K | \ll 1 \, .
\end{equation}
This expression has been obtained after neglecting small and slow spatial fluctuations in $ \, \rho_s(x) \, $ and
$ \, n_z(x) $. Besides, Eq.(\ref{DeltaFB}) is valid only provided that
\[
   T < \Delta({\cal E}) \le {\cal E} \ll E_B \, .
\]
The renormalization stops as soon as the energy $ \, {\cal E} \, $ becomes smaller than
$ \, \Delta \, $ since the fermions are massive. Accordingly, the effective fermionic gap
$ \, \Delta_{\rm f} \, $ can be determined from the self-consistency equation:
\begin{equation}\label{SelfCons}
    \Delta_{\rm f} = \Delta({\cal E})\Bigl|_{{\cal E} = \Delta_{\rm f}} \, .
\end{equation}
Equation (\ref{SelfCons}) leads to the result Eq.(\ref{Delta-L-Eff}) for
attractive, $ \, 1 < K < 2 $, or weakly repulsive, $ \, 0 < 1 - K \ll 1 $ interactions.
However, this straightforward way of calculations suffers from a serious
drawback since the renormalization procedure dynamically generates potentially
relevant multiparticle scatterings. It is technically difficult to
analyze their relevance within the functional bosonization approach.
%%%
% cannot be easily included into the consideration.
%%%
% The functional bosonization approach becomes cumbersome and technically difficult
% if one attempts to analyze a relevance of
% the multiplarticle scattering or to extend the theory beyond the weak interaction limit.
%%%
% Therefore, the above self-consistent derivation of the gap is incomplete.
%%%
To justify the self-consistent derivation of $ \, \Delta_{\rm f} \, $ and to
analyze the case of stronger repulsion, one needs an alternative approach.

Such an approach can be based on the bosonization of the fermionic part of
$ \, {\cal L} \, $ followed by the analysis of the fully bosonized Lagrangian:
\begin{equation}
\label{BosL}
   \tilde{\cal L}[\tilde{\phi}, \alpha ] \simeq {\cal L}_{\rm SG} +  \frac{v \left( \partial_x \alpha \right)^2}{8 \pi K} - i s \rho_s n_z \partial_\tau \alpha \, ;
\end{equation}
where at $ \, \epsilon=0 $
%%%
% [{\bf O.: I'd keep $ a $}]
%%%
\begin{eqnarray}
\label{SG-ferm}
%%%
% \frac{1}{4}{\cal L}_{\rm LL}(\alpha, K, v)
%%%
%                                  \frac{v}{8 \pi K} (\partial_x \alpha)^2
%%%
   {\cal L}_{\rm SG} = {\cal L}_{\rm LL}(\tilde \phi, K, v) +
        \frac{ \Delta_0 \sqrt{1 - n_z^2} }{2 \pi a}  \cos( 2 {\tilde \phi} ) ; & \\
%%%
% + \left[ \left( \epsilon e^{2 i ( {\tilde \phi} + \alpha)} + c.c. \right) \right] \!\! , \cr
%%%
%                             - i s \rho_s n_z \partial_\tau \alpha + \frac{v_F}{8 \pi} (\partial_x \alpha)^2 \, , \\
%%%
%   {\cal L}_{\rm LL} & \equiv & \frac{1}{2 \pi K v}
%                             \left[ \left( \partial_\tau {\tilde \phi} \right)^2 + \left( v \partial_x {\tilde \phi} \right)^2 \right] \! ;
%%%
   \label{GenLL}
   {\cal L}_{\rm LL}(\tilde{\phi}, K, v) \equiv \frac{1}{2 \pi K v}
                             \left[ \left( \partial_\tau \tilde{\phi} \right)^2 + \left( v \partial_x \tilde{\phi} \right)^2 \right] . &
\end{eqnarray}
%%%
% Here $ \, D_{\rm bs} \equiv L^{-1} \int_0^L {\rm d} x ( \Delta_0 \sqrt{1 - n_z^2} ) \, $ is the averaged backscattering amplitude,
%%%
Here $ \, a \sim v_F / E_B \, $ is the smallest spatial scale (the UV cut-off);
%%%
% $ \, v = v_F / K \, $ is the renormalized velocity of bosonic excitations;
%%%
$ \, \tilde{\phi} \, $ is a composite phase: $ \, {\tilde \phi} \equiv \phi - \alpha / 2 $,
with $ \, \phi \, $ being the usual bosonic phase -- its gradient is related to the fermionic
density: $ \, \partial_x \phi = - \pi (\rho_+ + \rho_- ) \, $ \cite{Giamarchi}.
%%%
% We note that the spatial fluctuations of $ \, \alpha \, $ are renormalized by the interactions [cf. the
% terms $ \, ( \partial_x \alpha )^2 \, $ in Eqs.(\ref{Lferm},\ref{BosL})].
%%%
% and the temporal ones are governed mainly by the cross-term $ \, O( n_z \partial_\tau) $
% [sub-leading time gradients are neglected].
%%%

Similar to Eq.(\ref{Lferm}), sub-leading gradients of $ \, \alpha \, $ are neglected in Eq.(\ref{BosL}).
%%%
% The scale separation allows us to substitute the averaged (over the coordinate)
% value $ \, {\bar D} \equiv L^{-1} \int_0^L {\rm d} x ( \Delta_0 \sqrt{1 - n_z^2} ) \, $ for $ \, \Delta_0
% \sqrt{1 - n_z^2} $.
%%
% Neglecting these small gradients is
%%%
% and using the averaged amplitude $ \, D_{\rm bs} \, $ are
%%%
This can be justified since $ \, {\cal L}_{\rm SG} \, $ corresponds to the quantum Sine-Gordon
model where the relevant vertex $ \, \sim \Delta_{0} \cos(2 \tilde{\phi}) \, $ generates
the bosonic mass at $ \, K < 2 \, $ \cite{IrrelVert} resulting in the scale separation between
bosonic and spin degrees of freedom.
%%%
% If $ \,
% K > 2 $,  this vertex becomes irrelevant because a strong attraction of the helical fermions leads to superconducting
% correlations which suppress the backscattering by the Kondo impurities.
%%%
% Parameters $ \, {\bar D} \, $ and $ \, K \, $ are renormalized in the Sine-Gordon theory. The renormalization in
% the underlying HLL is described by the Kosterlitz-Thouless equations, cf. \cite{MaciejkoLattice}. We need not
% the RG equations but the effective gap in the spectrum of bosons.
%%%
The effective gap $ \, \Delta \, $ in the spectrum of bosons can be determined with the help of
the Feynmann variational method \cite{Giamarchi}.
%%%
% , see Sect. E.2 in the book \cite{Giamarchi}.
%%%
After neglecting fluctuations of $ \, n_z $, the gap equation can be written as:
\begin{equation}\label{DeltaBos}
      \left(
         \frac{\Delta}{E_B} \right)^2 = \frac{ \Delta_0 }{E_B} \left( K \frac{\Delta}{E_B}
      \right)^K .
%%%
%      K < 2 \, .
%%%
\end{equation}
Equation (\ref{DeltaBos}) is valid at $ \, T < \Delta \, $ for $ \, K < 2 $, i.e., for arbitrary
strong repulsion as well as for weak and moderate attraction \cite{NoInelast}.
The multiparticle backscattering, which is described by (less relevant) higher
vertices $ \, \sim D^{(n)} \cos \left[ 2 n {\tilde \phi} \right] $ in $ \, {\cal L}_{\rm SG}$, $ \, n = 2, 3, \ldots $,
can be included into the gap equation. Such vertices yield corrections to the RHS of Eq.(\ref{DeltaBos}) of order
$ \, O( n^2 \times (D^{(n)} / E_B) \times ( K \Delta / E_B)^{n^2 K} ) $. Assuming that $ \, D^{(n)}
\ll E_B \, $ and $ \, \Delta \ll E_B $, contributions from the multiparticle backscattering are sub-leading and can
be neglected at $ \, K > 1/\log(E_B/\Delta_0) \, $ \cite{SublVert}.
%%%
% The higher vertices is less relevant and, if $ \, D_{n} \sim D $,
% they yield perturbatively small corrections to the gap which show up only at $ \, K \le 1/2 $. These
% corrections can be neglected without lost of generality.
%%%

%%%
% If quantum fluctuations of $ \, n_z \, $ are negligible,
%%%
At $ \, K \sim 1 $, Eqs.(\ref{SelfCons}) and (\ref{DeltaBos}) coincide --
the bosonic gap $ \, \Delta \, $ and the gap of the interacting helical fermions $ \, \Delta_{\rm f} \, $
are equivalent for weak and moderate interaction strengths. The
fluctuations of $ \, n_z \, $ result in small fluctuations of $ \, \Delta $. One can analyze a change of
the free energy caused by the gap fluctuations and prove an equivalence of the bosonic- and fermionic gaps
for arbitrary $ \, K $ \cite{BvsF-gaps}.

The effective low energy theory for the phase $ \, \alpha \, $ can now be derived either from the Lagrangian (\ref{Lferm})
or from its fully bosonized counterpart (\ref{BosL}). We use the former option,
namely, we restore the anisotropy, neglect $ \, g \, $ in the first term $ \, {\cal L} \, $ of (\ref{Lferm})
but take into account the interaction induced renormalizations, i.e., we use the effective gap  $ \,
{\Delta} $, the effective velocity $ \, v $, and the renormalized factor $ \, v / 8 \pi K \, $ in
the gradient term
%%%
% by using 1) the effective gap $ \, \Delta $, 2)~the renormalized velocity $ \, v $, and 3) the
% renormalized (inverse) amplitude of the spatial  fluctuations of $ \, \alpha $, $ \, v / 8 \pi K $,
%%%
rather than the bare values. We will use $ \, {\Delta} \, $ in further calculations since, unlike
$ \Delta_{\rm f} $, it allows us to describe the strong repulsion.

Integrating out the massive variables $ \, \{ \psi, \psi^\dagger
\} \, $ and $ \, n_z $, we obtain:
\begin{equation}\label{S-alpha-eps}
   {\cal L}_\alpha = {\cal L}_{\rm LL}(\alpha, K_\alpha, v_\alpha)
%%%
%      \frac{1}{2 \pi K_\alpha v_{\alpha}}
%         \Bigl(
%              (\partial_\tau \alpha)^2  +  (v_{\alpha} \partial_x \alpha)^2
%         \Bigr)
%%%
         - {\cal D} \,  \left( \epsilon e^{2 i \alpha} + c.c. \right) ;
\end{equation}
where we have defined
\begin{eqnarray}
    {\cal D} & \equiv & \frac{1}{4 \pi a} \frac{{\Delta}^2}{E_B} \log\left( \frac{E_B}{\Delta} \right) \! , \
                     a \, {\cal D} \ll {\Delta} ; \\
\label{Params-alpha}
    \frac{v_\alpha}{v} & = & \frac{K_\alpha}{4K} \simeq
%%%
% \frac{1}{\sqrt{\Upsilon}} , \ \Upsilon \equiv 1 +
%%%
    \frac{a \cal D}{ {\Delta} \sqrt{K} } \ll 1 .
\end{eqnarray}
At $ \, K = 1 $, Eqs.(\ref{S-alpha-eps}--\ref{Params-alpha}) are equivalent to the results of  Ref.\cite{AAY}.
The composite particles are slow in the absence of the interactions, $ \, v_\alpha(K = 1) / v_F \ll 1 $.
One can see that, as $ \, K \, $ decreases, the ratio $ \, v_\alpha / v \, $ increases remaining
small as long as $ \, K > 1 / \log(E_B/\Delta_0) $. Inequality $ \, v \gg v_\alpha \, $ reflects the scale separation
between the interacting (fast) helical fermions and the (slow) composite particles.
Eq.(\ref{S-alpha-eps}) is valid at $ \, K < 2 $, therefore $ \, K_\alpha \sim
a {\cal D} \sqrt{K} / {\Delta} $ is small with- and without interactions.

%%%
% the expression for $ \, \hat{H}_{f} \, $ reads as:
% \begin{equation}
% \label{HkF}
%      \hat{H}_{K}^{(f)}  = \int \!\! {\rm d} x \, \rho_s J_z S_z \left( \rho_{+} - \rho_{-} \right) , \
%             \rho_\pm \equiv \left( \psi^{\dagger} \psi \right)_{\pm} \, ; \\
% \end{equation}
% where $ \, J_z \, $ and $ \, S_z \, $ are the coupling constant and the spin operator.
%%%

%%%
% Our theory trivially includes the coupling
% $ \, J_z \, $ since one can always do a unitary transformation of the Hamiltonian and
% map the theory with $ \, J_z \ne 0 \, $ and given $ \, K \, $ to the case with  $ \, J_z = 0
% \, $ and the effective Luttinger parameter $ \, K_{\rm eff} = K + J_z \nu $,
%%%

%%%
% {\it DC transport in the HLL at $ \, \epsilon \ne 0 $}: If the electron-spin couplings is anisotropic,
% Eq.(\ref{S-alpha-eps}) is also the Sine-Gordon model.
%%%
{\it DC transport}: A regular anisotropy, $ \, \epsilon(x) = {\rm const} $,
would pin the phase, $ \, \alpha \simeq {\rm arg}(\epsilon) $. The pining is similar to the effect of
a magnetic field applied in the XY-plane: it breaks time reversal symmetry and opens a global gap in the
spectrum of the composite quasiparticles trivially blocking the ballistic dc transport.

Random fluctuations of $ \, {\rm arg}(\epsilon) $, Eq.(\ref{EpsFluct}), prevent
the global gap from opening but are able to localize the composite particles at
$ \, T \to 0 $. Indeed, Eq.(\ref{S-alpha-eps}) describes a disordered Sine-Gordon model
for which the localization was demonstrated long ago \cite{GiaSchulz}. Since $ \, K_\alpha
\ll 1 $, the localization length can be evaluated by the standard optimization procedure \cite{Giamarchi,Lloc}:
%%%
% : $ \, L_{\rm loc} \, $ is defined as a spatial sclae on which the typical energy governed by the last term in Eq.(\ref{S-alpha-eps})
% (i.e., the potential energy of the disorder) becomes equal to the energy governed by the term $ \, \propto
% (\partial_x \alpha)^2 \, $ in $ \, {\cal L}_{LL} $ (i.e., $ \, E_{\rm pin}$)
%%%

\begin{equation}
\label{L-loc}
   {L}_{\rm loc} \sim
%%%
% \frac{a}{ \left( w \log^{2}( E_B / \bar{\Delta} ) \right)^{1/3}} \left( \frac{E_B}{ K \bar{\Delta}} \right)^{4/3} \sim
%%%
                                 \frac{a}{ w^{1/3} }  \left( \frac{E_B}{ a {\cal D}  K^2 } \right)^{2/3};
%%%
%                                 K^{-1} ( v E_B / w D^2)^{1/3} ;
%%%
\end{equation}
$ w \, $ is defined in Eq.(\ref{EpsFluct}). Let us now discuss different temperature regimes of the dc transport.

The localization strongly effects the low temperature dc transport in not
too short samples, $ \, L \ge L_{\rm loc} $:
% sets three different regimes of transport:
% Localization has a vanishing effect on dc transport in short samples, $ \, L \ll
% L_{\rm loc} $. In the transient case,
%%%
the conductance is resistive (finite but smaller than the ballistic
one) and temperature-independent in transient samples with $ \, L \sim
L_{\rm loc} $, and the conductance vanishes in long samples, $ \, L \gg L_{\rm loc} $.

Sizable dc transport in long samples ($ L \gg L_{\rm loc} $) appears only at higher
temperatures: if $ \, T_{\rm MBL} < T < E_{\rm pin} $, the fermions are still gapped
but many-body states become delocalized and are able to support weak dc quantum transport
similar to transport in glassy systems. As $ \, T \to E_{\rm pin} $, the transport
becomes classical.
%%%
% If  $ \, T \le T_{\rm MBL} $ and the system is large, $ \, L \gg {L}_{\rm loc} $, dc transport is simply
% absent. In the opposite case $ \, L \ll {L}_{\rm loc} $, localization does not influence dc transport. In the
% transient case $ \, L \sim {L}_{\rm loc} \, $ the conductance is smaller than the ballistic one and
% temperature-independent until $ \, T \, $ reaches $ \, T_{\rm MBL} $. If $ \, T_{\rm MBL} < T < E_{\rm pin} $,
% the fermions are still gapped but many-body states of the composite particles become delocalized and
% they can support weak  dc  transport of a quantum nature. This transport is similar to that in glass-like
% systems at $ \, T_{\rm MBL} < T \ll E_{\rm pin} \, $ and it becomes classical (thermally activated) at $ \, T
% \to E_{\rm pin} $.
%%%
The straightforward estimate yields (cf. Ref.\cite{Giamarchi}):
\begin{equation}
\label{Epin}
   E_{\rm pin} \sim \frac{v_\alpha}{K_\alpha} L_{\rm loc}^{-1} \sim (w E_B)^{1/3} \left( \frac{a {\cal D}}{K} \right)^{2/3} .
%%%
% \left( w a {\cal D} / K \bar{\Delta} \right)^{1/3} \bar{\Delta} \ll \bar{\Delta} \, .
%%%
\end{equation}
The classical de-pinning energy is supposed to exceed the quantum energy scale, $ \, T_{\rm MBL} \ll E_{\rm pin} $
(a rigorous theory for $ \, T_{\rm MBL} \, $ at $ \, K_\alpha \ll 1 \, $ is missing). Semiclassical conductivity of
the composite particles can be estimated as $ \, \sigma_{\rm c} \propto v_\alpha K_\alpha \tau^{\rm (c)}_{\rm eff}
$ \cite{KuboSigma}, where $ \, \tau^{\rm (c)}_{\rm eff} \sim (T/E_{\rm pin} )^2 (K_\alpha E_{\rm pin})^{-1} \, $ is
an effective temperature dependent transport time \cite{TauEff}.
%%%
% The temperature dependence
% of $ \, \tau^{\rm (c)}_{\rm eff} \, $ results from the renormalization of the disorder, therefore $ \, \tau^{\rm (c)}_{\rm eff}
% \propto T^2 \, $ at $ \, K_\alpha \ll 1 $ \cite{Giamarchi}. If $ \, T \simeq E_{\rm pin} $, $ \, 1/\tau^{\rm (c)}_{\rm eff} \, $
% is of the order of energy governed by the term $ \, \propto (\partial_\tau \alpha)^2 \, $ in $ \, {\cal L}_{LL} $,
% Eq.(\ref{S-alpha-eps}) $ \, \Rightarrow \tau^{\rm (c)}_{\rm eff} \sim L_{\rm loc} / v_\alpha \sim 1 / K_\alpha E_{\rm pin} $.
%%%
Using Eq.(\ref{Params-alpha}), we arrive at
%%%
% is proportional to an effective transport time, $ \, \tau^{\rm (c)}_{\rm eff} $, and can be estimated
% as follows \cite{MityaOleg}:
%%%
\begin{equation}\label{Sigma-C}
%%%
%   E_{\rm pin} < T < \bar{\Delta}: \
%%%
   \sigma_{\rm c} \propto v_\alpha K_\alpha \tau^{\rm (c)}_{\rm eff}
            \sim \frac{v_F}{K_\alpha E_{\rm pin}} \left( \frac{K_\alpha}{K} \right)^2 \left( \frac{T}{E_{\rm pin}} \right)^2 \!\! .
\end{equation}
The factor $ ( K_\alpha/K )^2 \ll 1 \, $ reflects the suppressed Drude
weight. Equation (\ref{Sigma-C}) is valid provided that $ \, E_{\rm pin} < T \ll {\Delta} $.

Strictly speaking, Eq.(\ref{DeltaBos}) is valid at $ \, T < \Delta $. Nevertheless, we can draw some
qualitative conclusions for $ \, T \sim {\Delta} \ll E_B $, where the dc conductivity is dominated by
thermally activated fermionic quasiparticles and can be estimated as \cite{KuboSigma}:
\begin{equation}\label{Sigma-F}
%%%
%   \bar{\Delta} < T \ll E_B: \
%%%
   \sigma_{\rm f} \propto v K \tau^{\rm (f)}_{\rm eff}
            \sim v_F \tau_{0} \left( \frac{T}{E_B} \right)^{2(1-K)} \!\!\! ;
\end{equation}
where $ \, \tau_0 \, $ is governed by the disorder of the Kondo lattice, i.e., by the randomness
in $ \, \rho_s $. The theory for  $ \, \tau_0 \, $ is beyond the scope of the present Letter.
%%%
% Eq.(\ref{Sigma-F}) is valid in the range $ \, {\Delta} \le T \ll E_B $. At higher temperatures,
%%%
If $ \, \Delta \ll T \ll E_B $, the gap becomes temperature dependent and shrinks.
As a result, our theory looses its validity and the dc transport should reflect different physics.
%%%
% such that $ \, \xi \, $ overcomes the thermal length,
% the Kondo impurities become uncorrelated. It is known that an individual Kondo impurity has no effect
% on dc transport \cite{FurusakiMatveev}. Hence, the high-temperature conductance is expected to
% be ballistic.
%%%

{\it Validity}: Our consideration is based on several assumptions: Firstly, the Kondo array is
dense, the bare coupling constant is small and the XY-anisotropy is weak
\begin{equation}
\label{Params}
   \rho_s a \sim 1 , \quad \nu J_{\perp} \ll 1 , \quad |\epsilon|, w \ll 1 .
\end{equation}
Combining these inequalities with Eq.(\ref{Delta-L-Eff}), one can check
that $ \, {\Delta} \ll E_B \, $ and justify Eq.(\ref{SlowAlpha}).

%%%
% and confirms
% the scale separation between the fermionic subsystem and the composite particles.
%%%

Secondly, we have neglected the Kondo effect which is permissible only provided that
$ \, \Delta \, $ exceeds the Kondo temperature of a single Kondo impurity embedded into
the HLL, $ \, T_K $.
%%%
% , which does not show up in the dc conductance of the HLL with the single impurity \cite{FurusakiMatveev}, however,
% could completely change our conclusions due to suppression of backscattering in the Kondo array.
%%%
The standard estimate for $ \, T_K \, $ obtained from the RG approach reveals two regimes:
%%%
% is usually introduced as an energy scale at which the RG flow for the single impurity enters the strong coupling regime.
%%%
$ \, T^{(0)}_K \sim E_B \exp(-1/\nu J_{\perp}) \, $ is exponentially small at $ \, K \to 1 $ but
becomes larger due to the renormalization of $ \, J_{\perp} \, $ in the interacting case, $ \,
T_K^{\rm (int)} \sim E_B (\nu J_{\perp} / (1-K))^{1/(1-K)} \, $ at $ \, 1 - K \gg \nu J_{\perp} $ \cite{MaciejkoOregZhang,KondoTemp}.
%%%
% One can also obtain $ \, T_K^{\rm (int)} \, $ by considering a magnetization of the single spin-1/2 coupled
% to the interacting HLL and to an infinitesimal magnetic field $ \, \vec{h} = (0,0,h_z) $. $ \, T_K^{\rm (int)} \, $
% is the scale at which the perturbation theory in $ \, h_z \, $ fails to work.
%%%
Comparing $ \, T_K \, $ with $ \, \Delta \, $ from Eq.(\ref{Delta-L-Eff}), we find that $ \, T_K \, $
is always the smallest scale. Therefore, for a dense Kondo array, the Kondo screening can be neglected.
%%%
% ; see the inset in Fig.\ref{Delta_K}.
%%%

%%%
% Eqs.(\ref{Delta-L-Eff}) have been obtained
% for the case $ \, \nu J_{x,y}, | \epsilon | \ll 1 \, $ and they are valid if the coherence
% length is much larger than the smallest spatial scale (the UV cut-off):
% \begin{equation}\label{Valid-1}
%   \xi \gg a \sim v_F / E_B \, .
% \end{equation}
%%
% The gap reaches its maximum value at the moderate repulsion, see Fig.\ref{Delta_K}. Nevertheless,
% the condition (\ref{Valid-1}) always holds true in the case of the dense Kondo array, $ \, \rho_s a \le 1 $,
% which validates the theory.
%%%

Finally, taking into account restrictions on $ \, K $ discussed after Eq.(\ref{DeltaBos}),
we determine the validity range for the interaction strengths:
\begin{equation}
\label{K-lim}
    \bigl( \log\left( {E_B}/{\Delta_0} \right) \bigr)^{-1} < K < 2 \, .
\end{equation}
These conditions are not too restrictive, i.e., the presented theory is valid in the
broad range of interaction strength.

{\it Conclusions and open questions}: We have demonstrated that the
localization of the 1D helical electrons coupled to the random array
of the Kondo impurities is a generic and robust phenomenon which takes
place in the broad range of the electron interaction strengths, Eq.(\ref{K-lim}).
This confirms a qualitative conjecture of Ref.\cite{AAY}. We have found and
quantitatively described strong (non-perturbative) renormalizations of
physical parameters, Eq.(\ref{Delta-L-Eff}).
%%%
% Besides, we have determined the maximal strength
% of the repulsive interaction at which the theory is valid (this scale was
% not pointed out in Ref.\cite{AAY}).
%%%

We have discussed possible manifestations of localization in the
dc transport at low temperatures $ \, T \ll \Delta $ ($ \Delta \, $
is the gap in the electron spectrum caused by local spin ordering), when
the current is carried by slow composite spinon-fermion excitations.
A random anisotropy of the electron-spin coupling pins spin ordering
and localizes low-energy excitations.
The localization length $ \, L_{\rm loc} $ is given by Eq.(\ref{L-loc}).
The localization is known to lead to an insulating state, which persists
in a finite temperature interval
$ \, T < T_{\rm MBL} $, where $ \, T_{\rm MBL} \, $ is the temperature
of a many-body localization transition \cite{MBL}. $ \, T_{\rm MBL} \, $
is expected to be small compared to the classical de-pinning energy
$ \, E_{\rm pin} \ll \Delta $, Eq.(\ref{Epin}).
This leads to the following predictions for short samples, $ \, L \ll L_{\rm loc} $:
the deviation of the conductance from the universal ballistic value should be small
as $ \, L/L_{\rm loc} $. The low temperature transport of the samples with $ \, L \sim
L_{\rm loc} \, $ is characterized by strong mesoscopic fluctuations. Longer samples, $ \,
L \gg L_{\rm loc} $, are almost prefect insulators as long as $ \, T < T_{\rm MBL} $.
The theory for a temperature interval $ \, T_{\rm MBL} < T < E_{\rm pin} \, $ is yet
to be developed for both
long and short samples. At higher temperatures, $ \, E_{\rm pin} < T \ll \Delta $,
the dc transport becomes semiclassical with $ \, \sigma \sim
T^2 $, Eq.(\ref{Sigma-C}), and reduced Drude weight $ \, (K_\alpha / K)^2 \ll 1 $,
see Eq.(\ref{Params-alpha}) (and Eq.(3) in Ref.\cite{AAY}).
%%%
% These predictions can be exploited in experimental studies to
% identify the physical regimes for particular samples.
%%%

To develop the analytical theory, we had to restrict our
choice of parameters to those given in Eq.(\ref{Params}). In particular,
we have assumed a small anisotropy. This choice results in low $ \,
T_{\rm MBL} \, $ and large $ \, L_{\rm loc} $. However,
a strong XY-anisotropy looks more natural for the edge transport
and it could yield stronger localization with larger $ \, T_{\rm MBL}
\, $ and smaller $ \, L_{\rm loc} $. Therefore, we believe that the
considered mechanism of suppressing the ballistic transport can be
relevant for realistic samples. In particular, the temperature independent
transport observed in Ref.\cite{EdgeTransport-Exp2} might be associated
with the described above resistive regime of relatively short samples at
$ \, T <  T_{\rm MBL} $.

\begin{acknowledgments}
We gratefully acknowledge hospitality of the Abdus Salam ICTP where the part of this project was done.
V.I.Yu. also acknowledges hospitality of the Ludwig Maximilians University, Munchen,
and the partial support of RFBR grant 15-02-05657.
O.M.Ye. acknowledges support from the DFG through SFB TR-12, and the Cluster of Excellence, Nanosystems
Initiative Munich. We are grateful to Igor Aleiner, Alexander Nersesyan, and  Alexey Tsvelik for useful discussions.
\end{acknowledgments}

\bibliography{Bibliography}

\begin{thebibliography}{38}
\bibitem [{Irr()}]{IrrelVert}%
  \BibitemOpen
  \href@noop {} {}\bibinfo {note} {If $ \, K > 2 $, this vertex becomes
  irrelevant because a strong attraction of the helical fermions leads to
  superconducting correlations which suppress backscattering by the Kondo
  impurities.}\BibitemShut {Stop}%
\bibitem [{NoI()}]{NoInelast}%
  \BibitemOpen
  \href@noop {} {}\bibinfo {note} {We note that we do not consider interaction
  processes, which are sensitive to the phase space restriction and are similar
  to those discussed, for example, in
  Refs.\cite{GoldstGlazman-Short,GoldstGlazman-Long}.}\BibitemShut {Stop}%
\bibitem [{Sub()}]{SublVert}%
  \BibitemOpen
  \href@noop {} {}\bibinfo {note} {To ensure that all higher vertices yield
  only sub-leading corrections to the gap equation, we request $ \, (K \Delta /
  E_B)^K \ll 1 \, $ and solve this inequality at $ \, K \ll 1 $, where (with
  logarithmic accuracy) $ \, K^K \sim 1 \, $ and $ \, \Delta \sim
  \sqrt{\Delta_0 E_B} $, see Eq.(\ref{DeltaBos}). This results in the condition
  $ \, K \gg 1 / \log(E_B/\Delta_0) $.}\BibitemShut {Stop}%
\bibitem [{Llo()}]{Lloc}%
  \BibitemOpen
  \href@noop {} {}\bibinfo {note} {$ L_{\rm loc} \, $ is defined as a spatial
  scale on which the typical energy governed by the last term in
  Eq.(\ref{S-alpha-eps}) (i.e., the potential energy of the disorder) becomes
  equal to the energy governed by the term $ \, \propto (\partial_x \alpha)^2
  \, $ in $ \, {\cal L}_{\rm LL} $ (i.e., $ \, E_{\rm pin}$).}\BibitemShut
  {Stop}%
\bibitem [{Kub()}]{KuboSigma}%
  \BibitemOpen
  \href@noop {} {}\bibinfo {note} {The conductivity of a clean Luttinger Liquid
  with the Lagrangian (\ref{GenLL}) reads as $ \, \sigma(\omega) \propto i v K
  / \omega $ \cite{Giamarchi}. The estimate for $ \, \sigma_{c,f} \, $ is
  obtained from $ \sigma(\omega) \, $ after substituting $ \, v_\alpha K_\alpha
  \, $ (respectively $ \, v_F $) for $ \, v K \, $ and $ \, \omega + i /
  \tau^{(c,f)}_{\rm eff} \, $ for $ \, \omega \, $ at $ \, \omega \ll
  1/\tau^{(c,f)}_{\rm eff} $.}\BibitemShut {Stop}%
\bibitem [{Tau()}]{TauEff}%
  \BibitemOpen
  \href@noop {} {}\bibinfo {note} {If $ \, T \simeq E_{\rm pin} $,
  semiclassical $ \, \tau^{\rm (c)}_{\rm eff} \, $ is $ \, \propto L_{\rm loc}
  / v_\alpha \sim 1 / K_\alpha E_{\rm pin} $. This is the the scale on which
  terms $ \, \propto (\partial_\tau \alpha)^2 \, $ and $ \, \propto (v_\alpha
  \partial_x \alpha)^2 \, $ in $ \, {\cal L}_{\rm LL} $,
  Eq.(\ref{S-alpha-eps}), become equal. If $ \, T \ge E_{\rm pin} $, then the
  dc conductivity of a disordered Luttinger Liquid and, correspondingly, $ \,
  \tau^{\rm (c)}_{\rm eff} \, $ have a temperature dependence $ \, \propto
  T^{2(1-K_\alpha)} $, see Sect. 9.2 of the book \cite{Giamarchi} $ \,
  \Rightarrow \tau^{\rm (c)}_{\rm eff} \propto T^2 \, $ at $ \, K_\alpha \ll 1
  $. The temperature must be scaled by $ \, E_{\rm pin} \, $ to reproduce the
  value of $ \, \tau^{\rm (c)}_{\rm eff} \, $ at $ \, T \simeq E_{\rm pin} $.
  Combining all together we find the estimate for $ \, \tau^{\rm (c)}_{\rm eff}
  $.}\BibitemShut {Stop}%
\bibitem [{Kon()}]{KondoTemp}%
  \BibitemOpen
  \href@noop {} {}\bibinfo {note} {One can also obtain $ \, T_K^{\rm (int)} \,
  $ by considering a magnetization of the single spin-1/2 coupled to the
  interacting HLL and to an infinitesimal magnetic field $ \, \vec{h} =
  (0,0,h_z) $. $ \, T_K^{\rm (int)} \, $ is the scale on which the perturbation
  theory in $ \, h_z \, $ fails to work.}\BibitemShut {Stop}%
\end{thebibliography}

\end{document}